\documentclass[aps,prl,reprint,groupedaddress]{revtex4-1}
\usepackage{graphicx}
\begin{document}

%\preprint{}

\title{Solvable epidemic model on degree-correlated networks
}


\author{Satoru Morita}
\email[]{morita.satoru@shizuoka.ac.jp}
%\thanks{}
\affiliation{Department of Mathematical and Systems Engineering, Shizuoka University, Hamamatsu, 432-8561, Japan}


\date{\today}

\begin{abstract}
Disease and information spread over social and information networks.
Understanding the spread phenomena in networks requires
paying attention not only to the degree distribution but also to the degree correlation.
However, it is considered difficult to analytically deal with the effect of degree correlation on spread phenomena.
Here, we introduce degree correlation using a simple method and present the theoretical formulas
of the outbreak threshold and basic reproduction number.
We theoretically clarify the effect of the degree correlation.
\end{abstract}

%\keywords{}

\maketitle

\section{Introduction}
With the development of transportation and information transmission technology, 
diffusion processes on complex networks are attracting attention in many fields
\cite{barrad,fu,RMP2015}.
Outbreak of new infectious diseases that spread over human contact networks have 
threatened our lives, and computer viruses have through the internet 
caused severe economic damage worldwide. 
In considering the spread phenomena in networks,
one of the essential structures of the networks is the heterogeneity of degree $k$,
which is the number of connections each node has with other nodes
\cite{barrad,fu,newman2002,RMP2002,newman2006,RMP2015,barabasi2016}.
It is well known that many real networks have a scale-free property, whereby 
the degree distribution follows a power law for large values of $k$ \cite{RMP2002}, expressed as follows:
\begin{equation}
p_k\sim k^{-\gamma}.
\label{eq1}
\end{equation}
A remarkable feature of the scale-free networks is that 
the second moment $\langle k^2\rangle=\sum_k k^2p_k$
diverges when $\gamma \leq 3$.
However, some studies argue that many real networks are not strictly scale-free \cite{amaral2000,hamilton,broido2019}.
Even so, in most social networks, 
$\langle k^2\rangle$ is much larger than $\langle k\rangle^2$.
Such networks are collectively called fat-tailed networks \cite{barabasi2016}.

Another important structure is the degree correlation between two nodes connected by links.
The degree correlation is 
described using the degree correlation matrix $e_{kk'}$, which is 
the probability that one of the two ends of a randomly selected link 
has a node with degree $k$ and the other has a node with degree $k'$ \cite{newman2002,callaway,barabasi2016}.
This matrix is symmetric and expressed follows:
\begin{equation}
e_{kk'}=e_{k'k},
\label{eq01}
\end{equation}
and since $e_{kk'}$ is a probability, 
it is normalized as follows:
\begin{equation}
\sum_{k,k'}e_{kk'}=1.
\label{eq02}
\end{equation}
Because the probability that the node at the end of 
a randomly selected link has degree $k$ is always expressed as
\begin{equation}
q_k=\frac{kp_k}{\langle k\rangle},
\label{qk}
\end{equation}
the following relationship must be satisfied:
\begin{equation}
q_k=\sum_{k'}e_{kk'}.
\label{eq03}
\end{equation}
Moreover, the conditional probability that a node of degree $k$ is connected to 
a node of degree $k'$ is expressed as follows:
\begin{equation}
p(k'|k)=\frac{e_{kk'}}{q_k}.
\label{pkk'}
\end{equation}
If the network has no degree correlation, we obtain
\begin{equation}
e_{kk'}=q_k q_{k'},
\end{equation}
and thus $p(k'|k)=q_{k'}$.
The degree correlation coefficient is described using the Pearson correlation coefficient between
the degrees of two ends of the same link as follows:
\begin{equation}
r=\displaystyle \frac{\sum_{kk'}kk'(e_{kk'}-q_k q_{k'})}
{\sum_{k}k^2 q_k-(\sum_{k}k q_k)^2}
\end{equation}
Networks where $r>0$ are called assortative, whereas
networks where $r<0$ are caledl disassortative \cite{newman2002}.
Traditional social networks tend to be assortative, 
whereas online social networks tend to be disassortative \cite{hu2009}.

Epidemic models on static networks can be solved analytically using
the degree-based mean-field approximation when there is no degree correlation\cite{barrad,RMP2015,pastor2001a,pastor2001b,morita2016}. 
In this case, we obtain mathematical expressions using $\langle k \rangle$ and $\langle k^2 \rangle$
of the outbreak threshold $\lambda_c$ of the transmission rate,
above which prevalence can occur,
and the basic reproduction number $R_0$, which is the average number of secondary infections 
that a typical infection would directly cause in a completely susceptible population.
However, if there is degree correlation, we have no such analytical solution, 
although there is a formulation by the largest eigenvalue of the connectivity matrix or next-generation matrix
\cite{RMP2015,boguna2002,boguna2003}.
Therefore, the effect of correlations on the spread of infection remains unclear.
In this study, we propose a simple model that provides the degree correlation and derives an analytical solution
of the outbreak threshold $\lambda_c$ and basic reproduction number $R_0$.
We theoretically clarify that $R_0$ increases with the degree correlation.
As an extension, we derive an analytical solution in the case of bipartite population of men and women. 

\section{Model}

Here, to understand the effect of degree correlation, we use a simple formulation of $e_{kk'}$ that satisfies the above constraints
(Eqs.~(\ref{eq01}), (\ref{eq02}) and (\ref{eq03})) as follows:
\begin{equation}
e_{kk'}=q_k q_{k'}+\varepsilon p_k p_{k'}\left(\frac{k}{\langle k\rangle}-1\right)\left(\frac{k'}{\langle k\rangle}-1\right).
\label{eq6}
\end{equation}
Because $e_{kk'}\geq 0$, the tuning parameter $\varepsilon$ must be within the following range
\begin{equation}
-\frac{k_{\mbox{\tiny max}}^2}{(k_{\mbox{\tiny max}}-\langle k\rangle)^2}
<\varepsilon<\frac{k_{\mbox{\tiny max}}k_{\mbox{\tiny min}}}{(k_{\mbox{\tiny max}}-\langle k\rangle)(\langle k\rangle-k_{\mbox{\tiny min}})}.
\end{equation}
In the case of $\varepsilon>0$, the two nodes tend to connect ($e_{kk'}>q_k q_{k'}$)
when their degrees $k$ and $k'$ are both greater or lesser than the mean $\langle k\rangle$.
Therefore, if $\varepsilon>0$, the network is assortative, whereas
if $\varepsilon<0$, the network is dissassortative.
Indeed, the degree correlation coefficient $r$ is calculated as follows:
\begin{equation}
r=\varepsilon \frac{(\langle k^2\rangle-\langle k\rangle^2)^2}
{\langle k^3\rangle \langle k\rangle-\langle k^2\rangle^2}.
\label{r}
\end{equation}
As it can be easily proved, because 
$\langle k^3\rangle\langle k\rangle>\langle k^2\rangle^2$,
$r>0$ when $\varepsilon>0$, and $r<0$ when $\varepsilon<0$.
From Eq.~(\ref{pkk'}), the conditional probability is given as follows:
\begin{equation}
p(k'|k)=q_{k'}+\varepsilon p_{k'}\left(\frac{k'}{\langle k\rangle}-1\right)\left(1-\frac{\langle k\rangle}{k}\right),
\label{pkk'2}
\end{equation}
and the average nearest-neighbors' degree of nodes with degree $k$
is calculated as
\begin{equation}
\begin{array}{lcl}
k_{\mbox{\tiny nn}}(k)&=&\displaystyle \sum_{k'}k'p(k'|k)\\
&=&\displaystyle
\frac{\langle k^2\rangle}{\langle k\rangle}
+\varepsilon\frac{\langle k^2\rangle-\langle k\rangle^2}{\langle k\rangle}
\left(1-\frac{\langle k\rangle}{k}\right),
\label{annd}
\end{array}
\end{equation}
which is an increasing function of $k$ when $\varepsilon>0$. 

\section{Result}

\subsection{Simple epidemic model}
In this section, we consider an epidemic model presented in the works 
of Boguna et al. and Moreno et al. \cite{boguna2002,boguna2003,moreno},
which is expressed as follows:
\begin{equation}
 \frac{di_k(t)}{dt}=-i_k(t)+\lambda k
[1-i_k(t)]\sum_{k'}p(k'|k)i_{k'}(t),
\label{eq2}
\end{equation}
where $i_k(t)$ represents the density of infected
nodes  within each degree class $k$.
The first term on the right-hand side of Eq.~(\ref{eq2}) represents the recovery.
The second term represents the infection, which is proportional to the infection rate ($\lambda$), times the density of susceptible nodes ($1-i_k(t)$),  
the number of neighboring nodes ($k$), and the probability that any
neighbor is infected ($\sum_{k'}p(k'|k)i_{k'}(t)$).
Substituting Eq.~(\ref{eq6}) into Eq.~(\ref{eq2}), we obtain
\begin{equation}
\begin{array}{lcl}
\displaystyle \frac{di_k(t)}{dt}&=&-i_k(t)+\lambda (1-i_k(t))\\
&& \displaystyle\left[k\Theta(t)+ \varepsilon\left(k-\langle k\rangle\right)\left(\Theta(t)-I(t)\right)\right],
\end{array}
\label{eq2a}
\end{equation}
where $\Theta(t)$ represents the probability that an end of a randomly chosen link is infected,
which is expressed as follows:
\begin{equation}
\Theta(t)=\sum_{k} q_k i_{k}(t).
\label{Theta}
\end{equation}
$I(t)$ represents the fraction of infected nodes and is expressed as follows:
\begin{equation}
I(t)=\sum_{k} p_ki_{k}(t).
\label{rho}
\end{equation}
The equilibrium condition for Eq.~(\ref{eq2a}) results in
\begin{equation}
i_k^*=\frac{k\Theta^*+\varepsilon (k-\langle k\rangle)(\Theta^*-I^*)}
{1/\lambda +k\Theta^*+\varepsilon (k-\langle k\rangle)(\Theta^*-I^*)}.
\label{eq13}
\end{equation}
By substituting Eq. (\ref{eq13}) into Eqs.~(\ref{Theta}) and (\ref{rho}), we obtain the following
self-consistent equations 
\begin{equation}
\begin{array}{lcl}
\Theta^* &=& \displaystyle \sum_{k} \frac{k}{\langle k\rangle} p_k
\frac{k\Theta^*+\varepsilon (k-\langle k\rangle)(\Theta^*-I^*)}
{1/\lambda+ k\Theta^*+\varepsilon (k-\langle k\rangle)(\Theta^*-\rho^*)},\\
I^* &=& \displaystyle \sum_{k} p_k
\frac{k\Theta^*+\varepsilon (k-\langle k\rangle)(\Theta^*-I^*)}
{1/\lambda+ k\Theta^*+\varepsilon (k-\langle k\rangle)(\Theta^*-I^*)}.
\end{array}
\label{eq13}
\end{equation}
The self-consistent equations always have a zero solution $I^*=\theta^*=0$.
If the transmission rate $\lambda$ is above the outbreak threshold $\lambda_c$,
there is a nonzero solution,
which can be obtained through numerical calculation.
Fig.~1(a) shows an example of the calculation results,
which are similar to those demonstrated by Fig.~1 in the previous study \cite{newman2002}.
\begin{figure}
\begin{center}
\includegraphics[width=7cm]{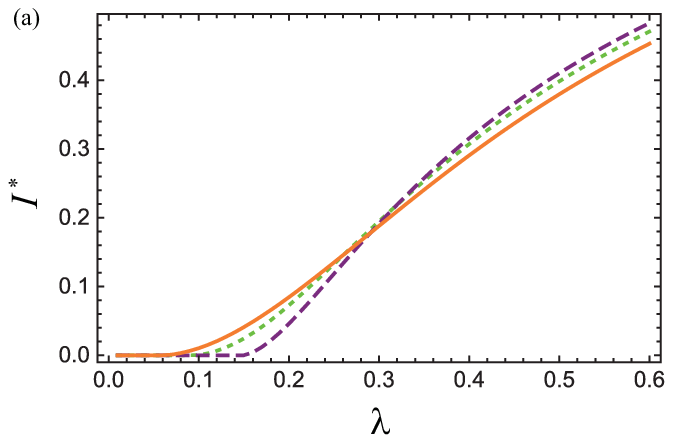}\\%
\includegraphics[width=7cm]{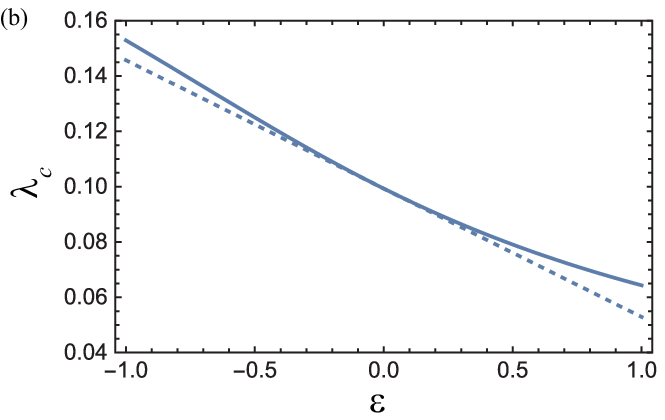}\\%
\includegraphics[width=7cm]{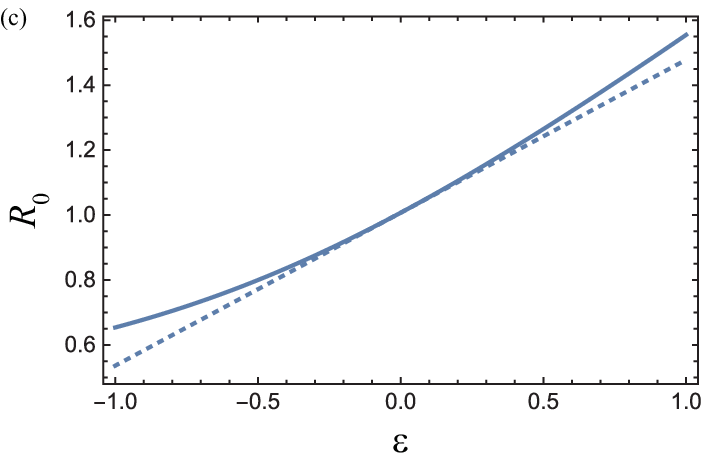}%
\caption{
(a) The fraction of infected nodes $I^*$ is plotted as a function of the trasmission rate 
$\lambda$ for some values of $\varepsilon$.
The curves are obtained by solving Eq.~({eq13}) numerically.
The orange (solid), green (dotted) and purple (dashed) curves correspond to an assortative network ($\varepsilon=1$),
an uncorrelated network ($\varepsilon=0$) and a disassortative network ($\varepsilon=-1$), respectively.
(b) The dependency of theoutbreak threshold $\lambda_c$ on $\varepsilon$.
The solid curve is given by Eq.~(\ref{lc}).
The dotted curve is obtained from an approximate expression (i.e. Eq.~(\ref{lcn})) that considers 
the first-order term of $\varepsilon$. 
(b) The dependency of the basic reproduction number $R_0$ on $\varepsilon$, when $\lambda=0.1$.
The solid and dotted curves represent Eq.~(\ref{r0}) and (\ref{r0n}), respectively.
Here, the degree distribution is $p_k\propto k^{-3}$ for $2\leq k \leq 1000$. 
\label{f1}}
\end{center}
\end{figure}

To calculate the outbreak threshold analytically, 
we consider the situation that the nonzero solution converges to 0. 
Taking the limit $I^*\to 0, \Theta^*\to 0$ of Eq.~(\ref{eq13}), we obtain
\begin{equation}
\begin{array}{lcl}
\Theta^* &=& \displaystyle \lambda_c  \left[\beta\Theta^*
+\varepsilon \alpha(\Theta^*-I^*)\right],\\
I^* &=& \displaystyle \lambda_c \langle k\rangle \Theta^*,
\end{array}
\label{eq14}
\end{equation}
where we set
\begin{equation}
\alpha=\frac{\langle k^2\rangle-\langle k\rangle^2}{\langle k\rangle}, \
\beta= \frac{\langle k^2\rangle}{\langle k\rangle}.
\end{equation}
By combining Eq.~(\ref{eq14}) and eliminating $I^*$ and $\theta^*$, we obtain
\begin{equation}
1=\lambda_c\left[\beta
+\varepsilon \alpha(1-\lambda_c \langle k\rangle)\right].
\label{eq17}
\end{equation}
As demonstrated in previous studies \cite{pastor2001a,pastor2001b},
if there is no degree correlation ($\varepsilon=0$), we obtain $\lambda_c= 1/\beta =\langle k\rangle/\langle k^2\rangle$.
Eq.~ (\ref{eq17}) has two solutions.
However, we select a solution that satisfies $\lambda_c= \langle k\rangle/\langle k^2\rangle$
for $\varepsilon=0$ 
\begin{equation}
\lambda_c  = \frac{\varepsilon\alpha+\beta-
\sqrt{(\varepsilon\alpha+\beta)^2-4\varepsilon\alpha\langle k\rangle}}
{2\varepsilon\alpha\langle k\rangle}.
\label{lc}
\end{equation}
By expanding Eq.~(\ref{lc}) around $\varepsilon=0$, we obtain
\begin{equation}
\lambda_c  =  \frac{\langle k\rangle}{\langle k^2\rangle} 
\left[1- \varepsilon \left(1-\frac{\langle k\rangle^2}{\langle k^2\rangle}\right)^2\right] + O(\varepsilon^2).
\label{lcn}
\end{equation}
Therefore, it is clear that the outbreak threshold $\lambda_c$ decreases when $\varepsilon$ increases (see Fig.~1(b)).

Next, we calculate the basic reproduction number $R_0$.
The Jacobi matrix $J_{kk'}$ of eq.~(\ref{eq2}) at the disease-free equilibrium ($i_k(t)=0$) is expressed as follows:
\begin{equation}
J_{kk'}= -\delta_{kk'}+\lambda k \sum_{k'}p(k'|k),
\label{jacobi}
\end{equation}
where $\delta_{kk'}$ is the Kronecker delta or identity matrix.
The second term of the right hand side of Eq.~(\ref{jacobi}) represents the transmission part.
In this case, 
the next-generation matrix of Eq.~(\ref{eq2}) is calculated as follows:
\begin{equation}
\begin{array}{lcl}
A_{kk'}&=&\lambda k p(k'|k)\\
&=&\displaystyle \lambda p_{k'}
\frac{kk'+\varepsilon \left(k'-\langle k\rangle\right)\left(k-\langle k\rangle\right)}{\langle k\rangle}.
\end{array}
\end{equation}
The basic reproduction number $R_0$ is determined by the spectral radius of the next-generation matrix
$A$.
Focusing on the following relations
\begin{equation}
\begin{array}{lcl}
\displaystyle \sum_{k'} q_{k'} A_{k'k}&=&\lambda (\beta+\varepsilon\alpha)q_k-\lambda \varepsilon \alpha p_k,\\
\displaystyle \sum_{k'} p_{k'} A_{k'k}&=&\lambda \langle k\rangle q_k,
\end{array}
\label{relation1}
\end{equation}
we see that the following vectors are the eigenvectors of $A$: 
\begin{equation}
2\langle k\rangle q_k-\left[\beta+\varepsilon\alpha\pm\sqrt{(\beta+\varepsilon\alpha)^2-4\varepsilon\alpha\langle k\rangle}\right]p_k.
\end{equation}
The corresponding eigenvalues are calculated as follows:
\begin{equation}
\lambda \langle k\rangle \frac{\beta+\varepsilon\alpha \mp \sqrt{(\beta+\varepsilon\alpha)^2-4\varepsilon\alpha\langle k\rangle}}{2}.
\label{evs}
\end{equation}
Because $A$ is a positive matrix, the Perron--Frobenius theorem shows that 
the solution with the plus sign before the route in Eq.~(\ref{evs}) is the dominant eigenvalue.
Therefore,
\begin{equation}
R_0=\lambda \langle k\rangle \frac{\beta+\varepsilon\alpha +\sqrt{(\beta+\varepsilon\alpha)^2-4\varepsilon\alpha\langle k\rangle}}{2}.
\label{r0}
\end{equation}
If we expand Eq.~(\ref{r0}) around $\varepsilon=0$, we obtain
\begin{equation}
R_0=\lambda \frac{\langle k^2\rangle}{\langle k\rangle}
\left[1+ \varepsilon \left(1-\frac{\langle k\rangle^2}{\langle k^2\rangle}\right)^2\right] + O(\varepsilon^2).
\label{r0n}
\end{equation}
Therefore, $R_0$ increases with the increase in $\varepsilon$ (see Fig.~1(c)).
By solving $R_0=1$ for Eq.~(\ref{r0}) and (\ref{r0n}), we obtain Eq.~(\ref{lc}) and (\ref{lcn}), respectively.

\subsection{Bipartite population}
We then consider then a population of men and women
who have the degree distributions $p_k^{(0)}$ and  $p_k^{(1)}$, respectively,
and assume that sexually transmitted diseases spread over the bipartite graph.
We use following expressions:
\begin{equation}
\begin{array}{clc}
\displaystyle \langle k \rangle_{(i)} &=&\displaystyle\sum_k k p_k^{(i)} \\
\displaystyle q_k^{(i)}&=&\displaystyle\frac{k}{\langle k \rangle_{(i)}}p_k^{(i)},
\end{array}
\end{equation}
for $i\in\{0,1\}$.
To consider the degree correlation, we expand Eq.~(\ref{eq6}) as follows:
\begin{equation}
e_{kk'}=q_k^{(0)} q_{k'}^{(1)}+\varepsilon p_k^{(0)} p_{k'}^{(1)}\left(\frac{k}{\langle k\rangle_{(0)}}-1\right)\left(\frac{k'}{\langle k\rangle_{(1)}}-1\right).
\label{eq6n}
\end{equation}
Note that $e_{kk'}$ is not symmetric generally, because the first subscript corresponds to men, 
and the second corresponds to women.
If $i_k^{(0)}(t)$ and  $i_k^{(1)}(t)$ are the densities of infected 
nodes within each degree class $k$ for men and women, respectively,
Eq.~(\ref{eq2}) is expanded as follows:
\begin{equation}
\begin{array}{lcl}
\displaystyle\frac{di_k^{(0)}(t)}{dt}&=&\displaystyle-i_k^{(0)}(t)+\lambda^{(0)} k
[1-i_k^{(0)}(t)]\sum_{k'}p^{(1)}(k'|k)i_{k'}^{(1)}(t),\\
\displaystyle\frac{di_k^{(1)}(t)}{dt}&=&\displaystyle-i_k^{(1)}(t)+\lambda^{(1)} k
[1-i_k^{(1)}(t)]\sum_{k'}p^{(0)}(k'|k)i_{k'}^{(0)}(t),
\end{array}
\end{equation}
where $\lambda^{(0)}$ represents the transmission rate from women to men, 
and $\lambda^{(1)}$ represent the inverse transmission rate.
Here, the conditional probability is expressed as follows: 
\begin{equation}
\begin{array}{lcl}
p^{(0)}(k'|k)&=&\displaystyle 
q_{k'}^{(0)}+\varepsilon p_{k'}^{(0)}\left(\frac{k'}{\langle k\rangle_{(0)}}-1\right)\left(1-\frac{\langle k\rangle_{(1)}}{k}\right),\\
p^{(1)}(k'|k)&=&\displaystyle 
q_{k'}^{(1)}+\varepsilon p_{k'}^{(1)}\left(\frac{k'}{\langle k\rangle_{(1)}}-1\right)\left(1-\frac{\langle k\rangle_{(0)}}{k}\right).
\end{array}
\label{pkk'2}
\end{equation}
In this case, the next-generation matrix is expressed as follows:
\begin{equation}
\left(
\begin{array}{cc}
0&A^{(01)}\\
A^{(10)}&0\end{array}
\right),
\label{ngm2}
\end{equation}
where the matrices $A^{(01)}$ and $A^{(10)}$ correspond to 
the transmissions from women to men and vice versa, respectively, 
and are given as 
\begin{equation}
\begin{array}{lcl}
A^{(01)}_{kk'}&=&\lambda^{(0)} k p^{(1)}(k'|k),\\
A^{(10)}_{kk'}&=&\lambda^{(1)} k p^{(0)}(k'|k).
\end{array}
\end{equation}
By using the following relation: 
\begin{equation}
\begin{array}{lcl}
\displaystyle \sum_{k'} q_{k'}^{(0)} A_{k'k}^{(01)}&=&\lambda^{(0)} (\beta^{(0)}+\varepsilon\alpha^{(0)})q_k^{(1)}
-\lambda^{(0)} \varepsilon \alpha^{(0)} p_k^{(1)},\\
\displaystyle \sum_{k'} p_{k'}^{(0)} A_{k'k}^{(01)}&=&\lambda^{(0)} \langle k\rangle_{(0)} q_k^{(1)},\\
\displaystyle \sum_{k'} q_{k'}^{(1)} A_{k'k}^{(10)}&=&\lambda^{(1)} (\beta^{(1)}+\varepsilon\alpha^{(1)})q_k^{(0)}
-\lambda^{(1)} \varepsilon \alpha^{(1)} p_k^{(0)},\\
\displaystyle \sum_{k'} p_{k'}^{(1)} A_{k'k}^{(10)}&=&\lambda^{(1)} \langle k\rangle_{(1)} q_k^{(0)},
\end{array}
\label{relation2}
\end{equation}
we can calculate the dominant eigenvalue of the next-generation matrix given by Eq.~(\ref{ngm2}).
When expanded to the first order, the basic reproduction number is written as follows:
%\begin{equation}
%\begin{array}{lcl}
%R_0^2&=& \displaystyle  \lambda^{(0)}\lambda^{(1)}
%\frac{\langle k^2\rangle_{(0)}}{\langle k\rangle_{(0)}}
%\frac{\langle k^2\rangle_{(1)}}{\langle k\rangle_{(1)}}\\
%& & \displaystyle +2\varepsilon  \lambda^{(0)}\lambda^{(1)}
%\left[\frac{\langle k^2\rangle_{(0)}-\langle k\rangle_{(0)}^2}{\langle k\rangle_{(0)}}\right]
%\left[\frac{\langle k^2\rangle_{(1)}-\langle k\rangle_{(1)}^2}{\langle k\rangle_{(1)}}\right]\\
%&& +O(\varepsilon^2).
%\end{array}
%\end{equation}
\begin{equation}
\begin{array}{lcl}
R_0&=& \displaystyle  \sqrt{\lambda^{(0)}\lambda^{(1)}
\frac{\langle k^2\rangle_{(0)}}{\langle k\rangle_{(0)}}
\frac{\langle k^2\rangle_{(1)}}{\langle k\rangle_{(1)}}}\\
& & \displaystyle \left[1+\varepsilon  
\left(1-\frac{\langle k\rangle_{(0)}^2}{\langle k^2\rangle_{(0)}}\right)
\left(1-\frac{\langle k\rangle_{(1)}^2}{\langle k^2\rangle_{(1)}}\right)\right]\\
&& +O(\varepsilon^2).
\label{r02}
\end{array}
\end{equation}
When $\varepsilon=0$, Eq.~(\ref{r02}) reproduces the well-known result demonstrated by May et al. \cite{may2001}.
Note that when the network composition is the same for men and women, 
Eq.~(\ref{r02}) coincides with Eq.~(\ref{r0n}) 

\section{Conclusion}

In this study, we introduce degree correlation using a simple method demonstrated in 
Eq.~(\ref{eq6})  and apply it to epidemic models.
As a result, we succeed in writing the outbreak threshold $\lambda_c$ and 
basic reproduction number $R_0$ using relatively simple mathematical formulas.
We theoretically clarify the effect of the degree correlation on $\lambda_c$ and $R_0$.
Parameter $\varepsilon$ introduced here is useful
because it is proportional to the degree correlation coefficient, $r$ (Eq.~(\ref{r})).
However, note that the degree correlation introduced here may not necessarily be realistic.
One of the deviations is the behavior of the average nearest-neighbors' degree $k_{\mbox{\tiny nn}}(k)$. 
According to Eq.~(\ref{annd}) it approaches constancy when $k$ increases.
However, there are reports that in many real networks it exhibits power law behavior $k_{\mbox{\tiny nn}}(k)\sim k^{\mu}$ \cite{barabasi2016}.
Eq.~(\ref{eq6}) can be expanded in various ways to obtain various forms of $k_{\mbox{\tiny nn}}(k)$.
However, the calculations performed here are too complicated to achieve analytical solutions.
In conclusion, our method is significantly useful in that it enables analytical handling.
We expect our simple formulations to be effective, especially in the case
where the maximum of degree $k$ in the network is not large because of some restrictions.
In real social networks, the links can change adaptively depending on the social conditions. 
The results in this study are expected to be a reference point when considering such dynamical networks.

\

\begin{acknowledgments}
This study was supported by the JSPS KAKENHI (no. 18K03453).
A part of this study was conducted at 
the Joint Usage / Research Center on Tropical Disease, Institute of Tropical Medicine, Nagasaki University (2020-Ippan-01), and at the Japan Science and Technology Agency Crest.
We would like to thank Editage (www.editage.com) for English language editing.
\end{acknowledgments}

\bibliography{ref2020.bib}

\end{document}